\documentclass[fleqn,twoside]{article}
\usepackage{espcrc2}                                                    
\usepackage{graphicx}        
\newcommand{\be}{\begin{equation}}
\newcommand{\ee}{\end{equation}}
\newcommand{\bea}{\begin{eqnarray}}
\newcommand{\eea}{\end{eqnarray}}

\title{Constraining the Dark Universe}
\author{Rachel Bean\address[IMPERIAL]{Theoretical Physics, The Blackett Laboratory, Imperial College, Prince Consort Road, London, U.K.},
Steen H. Hansen\address[OXFORD]{NAPL, University of Oxford,
Keble road, OX1 3RH, Oxford, UK}, Alessandro Melchiorri\addressmark[OXFORD]}

\begin{document}

\begin{abstract}
We combine complementary datasets to constrain dark energy.
Using standard Big Bang Nucleosynthesis and the observed abundances 
of primordial nuclides to put constraints on $\Omega_Q$
at temperatures near $T \sim 1MeV$, we find the strong constraint 
$\Omega_Q(\mbox{MeV}) < 0.045$ at $2\sigma$ c.l.. 
Under the assumption of flatness, using results from 
Cosmic Microwave Background (CMB) anisotropy measurements, 
high redshift supernovae (SN-Ia)
observations and data from local cluster abundances we put a new
constraint on the equation of state parameter 
$w_Q < -0.85$ at $68 \%$ c.l..
\end{abstract}

\maketitle

\bigskip

{\it Introduction.}
The discovery that the universe's evolution may be dominated by 
an effective cosmological constant  \cite{super1}, 
is one of the most remarkable cosmological findings of recent years. 
An exceptional opportunity is now opening up to decipher the nature of 
dark matter \cite{dark}, to test the veracity of theories and reconstruct 
the dark matters profile using a wide variety of observations 
over a broad redshift range.

One candidate that could possibly explain the observations is a
dynamical scalar ``quintessence'' field. One of the strong aspects of
quintessence theories is that they go some way explaining the
fine-tuning problem, why the energy density producing the acceleration
is $\sim 10^{-120}M_{pl}^{4}$. A vast range of ``tracker'' (see for
example \cite{quint,brax}) and ``scaling'' (for example
\cite{wett}-\cite{ferjoy}) quintessence models exist which approach
attractor solutions, giving the required energy density, independent
of initial conditions. The common characteristic of quintessence
models is that their equations of state, $w_{Q}=p/\rho$, vary with time
whilst a cosmological constant remains fixed at
$w_{Q=\Lambda}=-1$. Observationally distinguishing a time variation in
the equation of state or finding $w_Q$ different from $-1$ will
therefore be a success for the quintessential scenario.

We will here discuss observational constraints on general quintessence
models.

In \cite{bhm} we used standard big bang nucleosynthesis and the
observed abundances of primordial nuclides to put constraints on the
amplitude of the energy density, $\Omega_{Q}$, 
at temperatures near $T \sim 1 $MeV.
The inclusion of a scaling field increases the expansion rate
of the universe, and changes the ratio of neutrons to protons at
freeze-out and hence the predicted abundances of light elements. 

In \cite{rachel} we have then combined the latest observations of the 
Cosmic Microwave Background (CMB) anisotropies provided by
the Boomerang \cite{Boom2}, DASI \cite{Dasi} and Maxima \cite{Maxima} 
experiments and the information from Large
Scale Structure (LSS) with the luminosity distance of high 
redshift supernovae (SN-Ia) to put constraints on the dark energy 
equation of state parameterized by a redshift independent 
quintessence-field pressure-to-density ratio $w_Q$.
We also made use of the Hubble Space Telescope (HST) constraint
on the Hubble parameter $h=0.72 \pm 0.08$ \cite{freedman}. 

We will briefly review the results obtained in those paper in the
next sections.

\medskip
{\it Early-Universe Constraints from BBN.}
In the last few years important experimental progress
has been made in the measurement of light element primordial abundances.
For the $^4He$ mass fraction, $Y_{\mbox{He}}$, two marginally compatible
measurements have been obtained from regression against zero metallicity
in blue compact galaxies. A low value  $Y_{\mbox{He}} =
0.234 \pm 0.003$ \cite{helow} and a high one $Y_{\mbox{He}} =
0.244 \pm 0.002$ \cite{hehigh} give realistic bounds.
We use the high value in our analysis; if one instead considered
the low value, the bounds obtained would be even stronger.

Observations in different quasar absorption line
systems give a relative abundance of  deuterium, which is 
critical in fixing the
baryon fraction, of $D/H=(3.0 \pm 0.4) \cdot 10^{-5}$
\cite{delow}.                                                                 

\begin{figure}[htb]
\begin{center}
\includegraphics[scale=0.35]{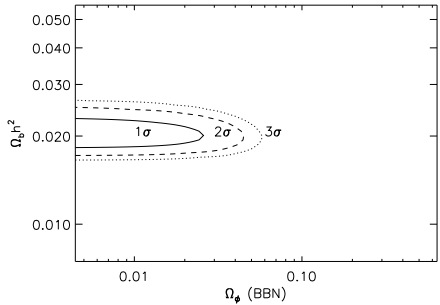}
\end{center}
\caption{$1,2$ and $3 \sigma$ likelihood contours in the $(\Omega_b
h^2, \Omega_Q (1 \mbox{MeV}))$ parameter space derived from $^4He$
and $D$ abundances.}
\label{figbbn}
\end{figure}

In the standard BBN scenario, the primordial abundances are a function
of the baryon density $\eta \sim \Omega_bh^2$ only. In order to put
constraints on the energy density of a primordial field a $T\sim$ MeV,
we modified the standard BBN code ~\cite{kawano} by including the
quintessence energy component $\Omega_Q$.
We then performed a likelihood analysis in
the parameter space $(\Omega_bh^2,\Omega_{Q}^{BBN})$
using the observed abundances $Y_{\mbox{He}}$ and $D/H$.
In Fig.~\ref{figbbn} we plot the $1,2$ and $3 \sigma$ likelihood
contours in the $(\Omega_b h^2, \Omega_Q^{BBN})$ plane.
 
Our main result is that the experimental data for $^4He$ and $D$ does
not favour the presence of a dark energy component, providing the
strong constraint $\Omega_Q(\mbox{MeV}) < 0.045$ at $2\sigma$
(corresponding to $\lambda > 9$ for the exponential potential
scenario), strengthening significantly the previous limit
of ~\cite{Ferreira:1998hj,ferjoy} $\Omega_Q(\mbox{MeV}) < 0.2$.
The reason for the difference is due to
the improvement in the measurements of the observed abundances,
especially for the deuterium, which
now corresponds to approximately $\Delta N_{\mbox{eff}} < 0.2-0.3$
additional effective neutrinos (see, e.g. \cite{burles}),
whereas Ref.~\cite{Ferreira:1998hj,ferjoy}
used the conservative value $\Delta N_{\mbox{eff}} < 1.5$.
One could worry about the effect of any underestimated systematic
errors, and we therefore multiplied the error-bars of the observed
abundances by a factor of $2$. Even taking this into account, there is
still a strong constraint $\Omega_Q(\mbox{MeV}) < 0.09$ ($\lambda
>6.5$) at $2\sigma$.
                        
\medskip

{\it Constraints on the Dark energy equation of state.}

The importance of combining different data sets in order to obtain
reliable constraints on $w_Q$ has been stressed by
many authors (see e.g. \cite{PTW}, \cite{hugen},\cite{jochen}), 
since each dataset suffers from degeneracies between the various
cosmological parameters and $w_Q$ . Even if one restricts consideration
 to flat universes and to a value of $w_Q$ constant
in time then the SN-Ia luminosity distance and position of the
first CMB peak are highly degenerate in $w_Q$ and $\Omega_Q$,
the energy density in quintessence.

The effects of varying $w_Q$ on the angular power spectrum of the
CMB anisotropies can be reduced to just two . 
Firstly, since the inclusion of quintessence
changes the overall content of matter and energy, the angular
diameter distance of the acoustic horizon size at recombination will be altered. 
In flat models (i.e. where the energy density in matter is
equal to $\Omega_M=1-\Omega_Q$), 
this creates a shift in the
peaks positions of the angular spectrum as 
\begin{eqnarray}{\cal R}&=&\sqrt{(1-\Omega_Q)}y, \label{Req} \\
y&=&\int_0^{z_{dec}}
[(1-\Omega_Q)(1+z)^3 \nonumber \\
&&+\Omega_{Q}(1+z)^{3(1+w_Q)}]^{-1/2} dz.
\nonumber \end{eqnarray}
It is important to note that the effect is completely degenerate
in the interplay between $w_Q$ and $\Omega_Q$.
Furthermore, it does not add any new
features beyond those produced by the presence of a
cosmological constant \cite{eb}, and it is not particularly sensitive
to further time dependencies of $w_Q$.

\begin{figure}[thb]
\begin{center}
\includegraphics[scale=0.35]{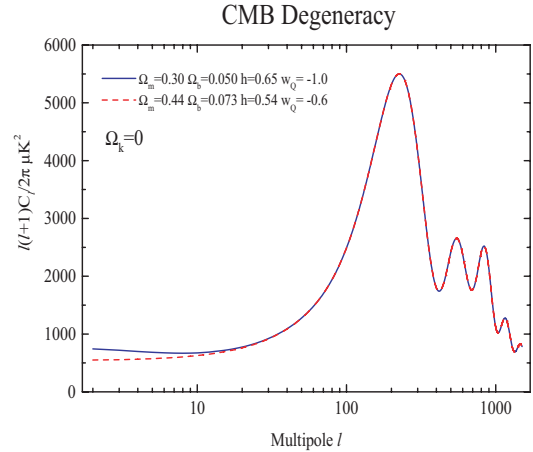}
\end{center}
\caption{CMB power spectra and the angular diameter distance 
degeneracy. The models are computed assuming flatness, 
$\Omega_k=1-\Omega_M-\Omega_Q=0$). The Integrated Sachs Wolfe effect on large angular scale slightly breaks the degeneracy. The degeneracy can be broken with a strong prior on $h$, in this paper we use the results from the HST.}
\label{figomega}
\end{figure}

Secondly, the time-varying Newtonian potential after decoupling will
produce anisotropies at large angular scales through the Integrated
Sachs-Wolfe (ISW) effect. The curve in the CMB angular spectrum
on large angular scales depends not only on the value of
$w_Q$ but also its variation with redshift.
However, this effect will be difficult to
disentangle from the same effect generated by a cosmological constant,
especially in view of the affect of cosmic variance and/or
gravity waves on the large scale anisotropies.

In order to emphasize the importance of degeneracies between
all these parameters while analyzing the CMB data, we plot
in Figure 2 some degenerate spectra, obtained keeping
the physical density in matter $\Omega_Mh^2$, the physical density
in baryons $\Omega_bh^2$ and
${\cal R}$ fixed. As we can see from the plot, models degenerate
in $w_Q$ can easily be constructed.
However the combination of CMB data with other 
different datasets can break the mentioned degeneracies.

\begin{figure}[thb]
\begin{center}
\includegraphics[scale=0.35]{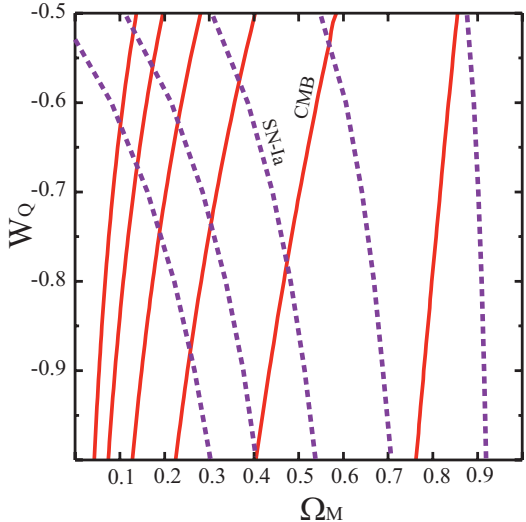}
\end{center}
\caption{Contours of constant ${\cal R}$ (CMB) and 
$SN-Ia$ luminosity distance in the $w_Q$-$\Omega_M$ plane.
The degeneracy between the two distance measures can be broken 
 by combining the two sets of complementary information. 
The luminosity distance is chosen to be equal to $d_{l}$ at
$z=1$ for a fiducial model with 
$\Omega_{\Lambda}=0.7$, $\Omega_{M}=0.3$,$h=0.65$.
(We note that as $\Omega_{Q}=1-\Omega_M$ 
goes to zero the dependence of ${\cal R}$ and $d_{L}$ upon 
$w_{Q}$ also become zero, as there is no dark energy present.)
Figure taken from \cite{rachel}}
\label{figomega2}
\end{figure}
\medskip

Type Ia supernovae, in particular, can be extremely useful in this.
Evidence that the universe's expansion rate was accelerating was first 
provided by two groups, the SCP and High-Z Search Team(\cite{super1}) using 
type Ia supernovae (SN-Ia) to probe the nearby expansion dynamics. 
SN-Ia seem to be 
good standard candles, as they exhibit a strong phenomenological 
correlation between the decline rate and peak magnitude of the 
luminosity. The observed apparent bolometric luminosity 
is related to the luminosity distance, measured in Mpc, by
$m_{B}=M+5 log d_{L}(z)+25$,
where M is the absolute bolometric magnitude. 
The luminosity distance is sensitive to the cosmological 
evolution through an integral dependence on the Hubble factor 
$d_{l}=(1+z)\int_{0}^{z} (dz'/H(z',\Omega_{Q},w_{q})$,
and can therefore be used to constrain the scalar equation of state. 
As can be seen in Figure \ref{figomega2} 
there is an inherent degeneracy in the luminosity distance in the 
$\Omega_{M}/w_{Q}$ plane; one can see that little can be said
about the equation of state from luminosity distance data alone. However,
 the degeneracies of CMB and SN1a data complement 
one another so that together they offer a more powerful approach 
for constraining $w_Q$. 

\medskip
\begin{table}[bt]
\renewcommand*{\arraystretch}{1.5}
\caption{
  Constraints on $w_Q$ and $\Omega_M=1-\Omega_Q$ 
  using different priors and datasets.
  We always assume flatness and $t_0>10$~Gyr.
  The $1\sigma$ limits are found from the 16\% 
  and 84\% integrals of the marginalized likelihood. 
  The HST prior is $h=0.72 \pm0.08$ while for the BBN prior
  we use the conservative bound $\Omega_bh^2=0.020\pm0.005$.
}\label{tab}
\begin{tabular}{lrrrr}
\hline
\\
CMB+HST
&$w_Q<-0.62$
\\
&$0.15<\Omega_M<0.45$
\\
CMB+HST+BBN
&$-0.95<w_Q<-0.62$
\\
&$0.15<\Omega_M<0.42$
\\
CMB+HST+SN-Ia
&$-0.94<w_Q<-0.74$
\\
&$0.16<\Omega_M<0.34$
\\
CMB+HST+SN-Ia+LSS
&$w_Q<-0.85$
\\
&$0.28<\Omega_M<0.43$
\\
\hline
\\
\end{tabular}
\end{table}

\begin{figure}[thb]
\begin{center}
\includegraphics[scale=0.35]{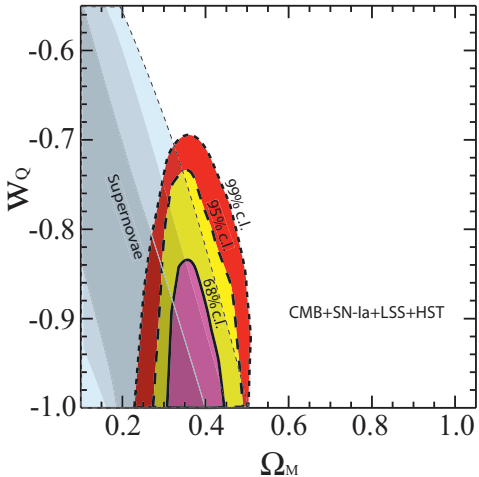}
\end{center}
\caption{The likelihood contours in the ($\Omega_M$, $w_Q$) plane,
with the remaining parameters taking their best-fitting values for the
joint CMB+SN-Ia+LSS analysis described in the text.
The contours correspond to 0.32, 0.05 and 0.01 of the peak value of the
likelihood, which are the 68\%, 95\% and 99\% confidence levels respectively.}
\label{figo}
\end{figure}
\medskip

Table I shows the $1$-$\sigma$ constraints on $w_Q$ for different 
combinations of priors, obtained in \cite {rachel} after marginalizing
over all remaining {\it nuisance} parameters.
The analysis is restricted to {\it flat} universes.
One can see that $w_Q$ is poorly constrained from CMB data alone,
even when the strong HST prior on the Hubble parameter, 
$h=0.72\pm0.08$, is assumed.
Adding a Big Bang Nucleosynthesis prior, 
$\Omega_bh^2 =0.020 \pm 0.005$, has small effect on the CMB+HST 
result.
Adding SN-Ia breaks the CMB $\Omega_Q-w_Q$ degeneracy and 
improves the upper limit on $w_Q$, with $w_Q <-0.74$.
Finally, including information  from local cluster abundances through 
$\sigma_8=(0.55\pm0.1)\Omega_M^{-0.5}$, where $\sigma_8$ is the {\it rms} 
mass fluctuation in spheres of $8 h^{-1}$ Mpc, further breaks
the quintessential-degeneracy, giving $w_Q <-0.85$ at $1$-$\sigma$.
Also reported in Table I, are the constraints on $\Omega_M$.
As we can see, the combined data suggests the presence of dark
energy with high significance, even in the case CMB+HST.
It is interesting to project our likelihood in the $\Omega_Q-w_Q$ plane. 
Proceeding as in \cite{melk2k}, we attribute a likelihood to a point
in the ($\Omega_M$, $w_Q$) plane by finding the remaining parameters that
maximize it. We then define our $68\%$, $95\%$
 and  $99\%$ contours to be where the likelihood falls to $0.32$, $0.05$ and
$0.01$ of its peak value, as would be the case for a
two dimensional multivariate Gaussian.
In Figure 4 we plot likelihood contours in the
($\Omega_M$, $w_Q$) plane for the joint analyses of
CMB+SN-Ia+HST+LSS data together with the contours from the SN-Ia dataset
only. As we can see, the combination of the data breaks the
luminosity distance degeneracy.
\medskip

\medskip
{\it Conclusions.}

We have provided new constraints on the dark energy 
by combining different cosmological data.
We have examined BBN abundances in a cosmological
scenario with a scaling field.  We have quantitatively discussed how
large values of the fractional density in the scaling field
$\Omega_{Q}$ at $T \sim 1 $MeV can be in agreement with the
observed values of $^4He$ and $D$, assuming standard Big Bang
Nucleosynthesis.  The $2\sigma$ limit $\Omega_{Q}(1 \mbox{MeV}) <
0.045$ severely constrains a wide class of quintessential scenarios,
like those based on an exponential potential.
For example, for the pure exponential
potential the total energy today is restricted to
$\Omega_{Q}={3 \over 4} \Omega_{Q}(1 \mbox{MeV}) \le0.04$. 
This result put strong constraints on the presence of quintessence
during recombination.

We have then provided constraints on the equation of state parameter
$w_Q$. 
The new CMB results provided by Boomerang and DASI improve the constraints 
from previous and similar analysis (see e.g., \cite{PTW},\cite{bondq}), with
$w_Q<-0.85$ at $68 \%$ c.l. ($w_Q<-0.76$ at $95 \%$ c.l.).
We have also demonstrated how the combination of CMB data 
with other datasets is crucial in order to break 
the $\Omega_Q-w_Q$ degeneracy.
The constraints from each single datasets are, as expected, 
quite broad but compatible with each other, providing an important 
consistency test.
When comparison is possible (i.e. restricting to similar priors and
datasets), our analysis is compatible with other recent analysis
on $w_Q$ (\cite{others}).
Our final result is perfectly in agreement with the $w_Q=-1$ 
cosmological constant case and gives no support to a 
quintessential field scenario with $w_Q > -1$.
A frustrated network of domain walls or
a purely exponential scaling field are excluded at high 
significance. In addition a number of quintessential models 
are highly disfavored, e.g. power law potentials with 
$p\ge 1$ and the oscillatory 
potential, to name a few.

It will be the duty of higher redshift datasets, for example
from clustering observations \cite{calvao} to point to 
a variation in $w$ that might place quintessence in a more 
favorable light. 

The result obtained here, however, could be plagued by some
of the theoretical assumptions we made.
The CMB and LSS constraints can be weakened by the inclusion of 
additional relativistic degrees of freedom~\cite{Bowen:2001in}, by a 
background of gravity waves, of isocurvature perturbations or
by adding features in the primordial perturbation spectra.
These modifications are not expected in the most basic 
and simplified inflationary scenario but they are still 
compatible with the present data.
The SN-Ia result has been obtained under the assumption of a 
constant-with-time $w_Q>-1$.
Inclusion of the region with $w_Q<-1$ could affect
our constraints (\cite{maor2}). 
In \cite{rachel} we have shown that in general $w_{eff}$ is
a rather good approximation for dynamical quintessential models 
since the luminosity distance depends on
$w_Q$ through a multiple integral that smears its redshift 
dependence, and our results are
therefore valid for a wide class of quintessential models.
This `numbing' of sensitivity to $w_Q$, first noticed by
\cite{maor},  implies that maybe 
an effective equation of state is the most tangible parameter 
able to be extracted from supernovae. However with the promise of 
large data sets from Planck and SNAP satellites, 
opportunities may yet still be open to reconstruct a time varying 
equation of state \cite{jochen}.

\medskip

\textit{Acknowledgements} It is a pleasure to thank 
Ruth Durrer, Pedro Ferreira, Massimo Hansen and Matts Roos
for comments and suggestions. RB and AM are supported by PPARC. 
SHH is supported by a Marie Curie Fellowship of the 
European Community under the contract HPMFCT-2000-00607. 
We acknowledge the use of CMBFAST~\cite{CMBFAST}.

\end{document}